\documentclass[a4paper,11pt]{article}
\usepackage{pos}

\usepackage{subcaption}
\usepackage{cleveref}
\usepackage{verbatim}
\title{A prototype tank for the SWGO detector}

\author*{Sofia Grusovin}
\author[a,f]{Giovanni Consolati}
\author[b,e]{Alessandro de Angelis}
\author[e]{Cornelia Arcaro}
\author[c,g]{Francesca Bisconti}
\author[c,g]{Andrea Chiavassa}
\author[b,e]{Michele Doro}
\author[d,h]{Fausto Guarino}
\author[b,e]{Mosè Mariotti}
\author[b,e]{Elisa Prandini}

\affiliation[a]{Department of Aerospace Science and Technology, Politecnico di Milano,\\
Via LaMasa 34, 20156 Milano, Italy}

\affiliation[b]{Dipartimento di Fisica e Astronomia G. Galilei, Università degli Studi di Padova,\\
Via F. Marzolo 8, 35131 Padova, Italy}

\affiliation[c]{Dipartimento di Fisica, Università degli Studi di Torino,\\
Via Pietro Giuria 1, 10125 Torino, Italy}

\affiliation[d]{Dipartimento di Fisica "Ettore Pancini", Università degli Studi di Napoli Federico II,\\
Strada Comunale Cinthia, 80126 Napoli, Italy}

\affiliation[e]{INFN Padova,\\
Via Francesco Marzolo 8, 35131 Padova, Italy}

\affiliation[f]{INFN Milano,\\
Via Celoria 16, 20133 Milano, Italy}

\affiliation[g]{INFN Torino,\\
Via Pietro Giuria 1, 10125 Torino, Italy}

\affiliation[h]{INFN Napoli,\\
Strada Comunale Cinthia, 80126 Napoli, Italy}

\emailAdd{sofia.grusovin@protonmail.com}
\emailAdd{giovanni.consolati@polimi.it}
\emailAdd{alessandro.deangelis@unipd.it}
\emailAdd{cornelia.arcaro@pd.infn.it}
\emailAdd{fbiscont@to.infn.it}
\emailAdd{achiavas@to.infn.it}
\emailAdd{michele.doro@unipd.it}
\emailAdd{guarino@na.infn.it}
\emailAdd{mose.mariotti@unipd.it}
\emailAdd{elisa.prandini@unipd.it}

\abstract{ The Southern Wide-field Gamma-ray Observatory (SWGO) is an international collaboration working on realizing a next-generation observatory located in the Southern hemisphere, which offers a privileged view of our galactic center. We are working on the construction of a prototype water Cherenkov detector at Politecnico di Milano using a flexible testing facility for several candidate light sensors and configurations.\\
A structure able to hold different types of detectors in multiple configurations has been designed, built and tested in Politecnico’s labs. Furthermore, an analytical study of muons and electrons showers has been carried out using the SWGO observatory simulation software to examine the correlation between the detection capabilities of the prototype tank and its water level.}

\FullConference{%
  *** 27th European Cosmic Ray Symposium - ECRS ***\\
  *** 25-29 July 2022 ***\\
  *** Nijmegen, the Netherlands ***
}

\begin{document}
\maketitle

\section{Introduction}
\begin{figure}[h!]
    \centering
    \subfloat[Qualitative representation of SWGO’s layout.\cite{loi}\label{fig:array1}]{
        \includegraphics[height=4cm]{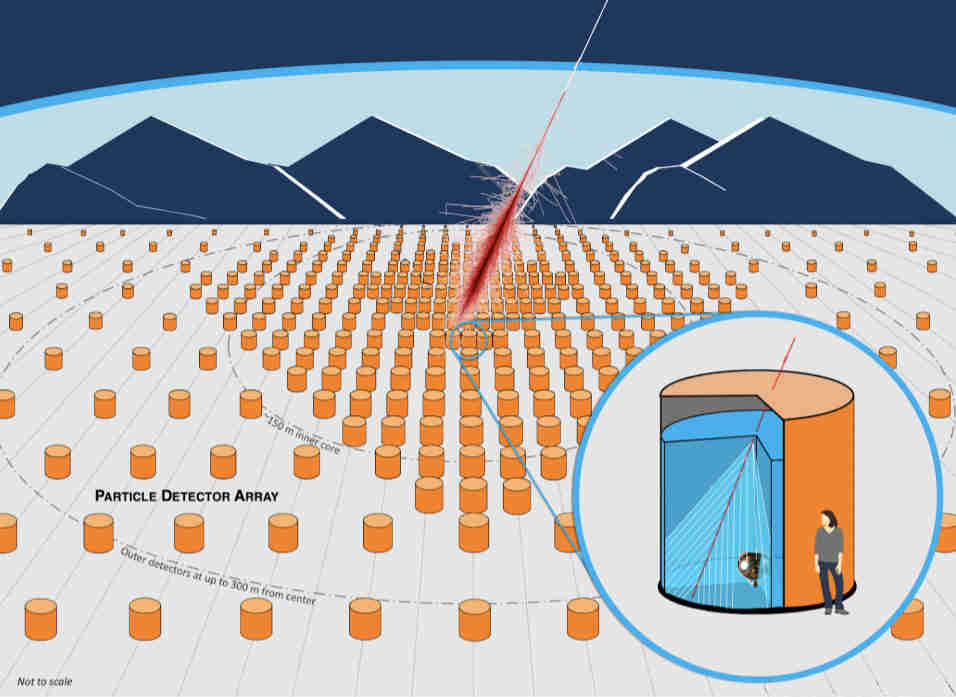}
    }
    \quad
    \subfloat[FoV of SWGO and HAWC.\cite{loi}\label{fig:fov}]{
        \includegraphics[height=4cm]{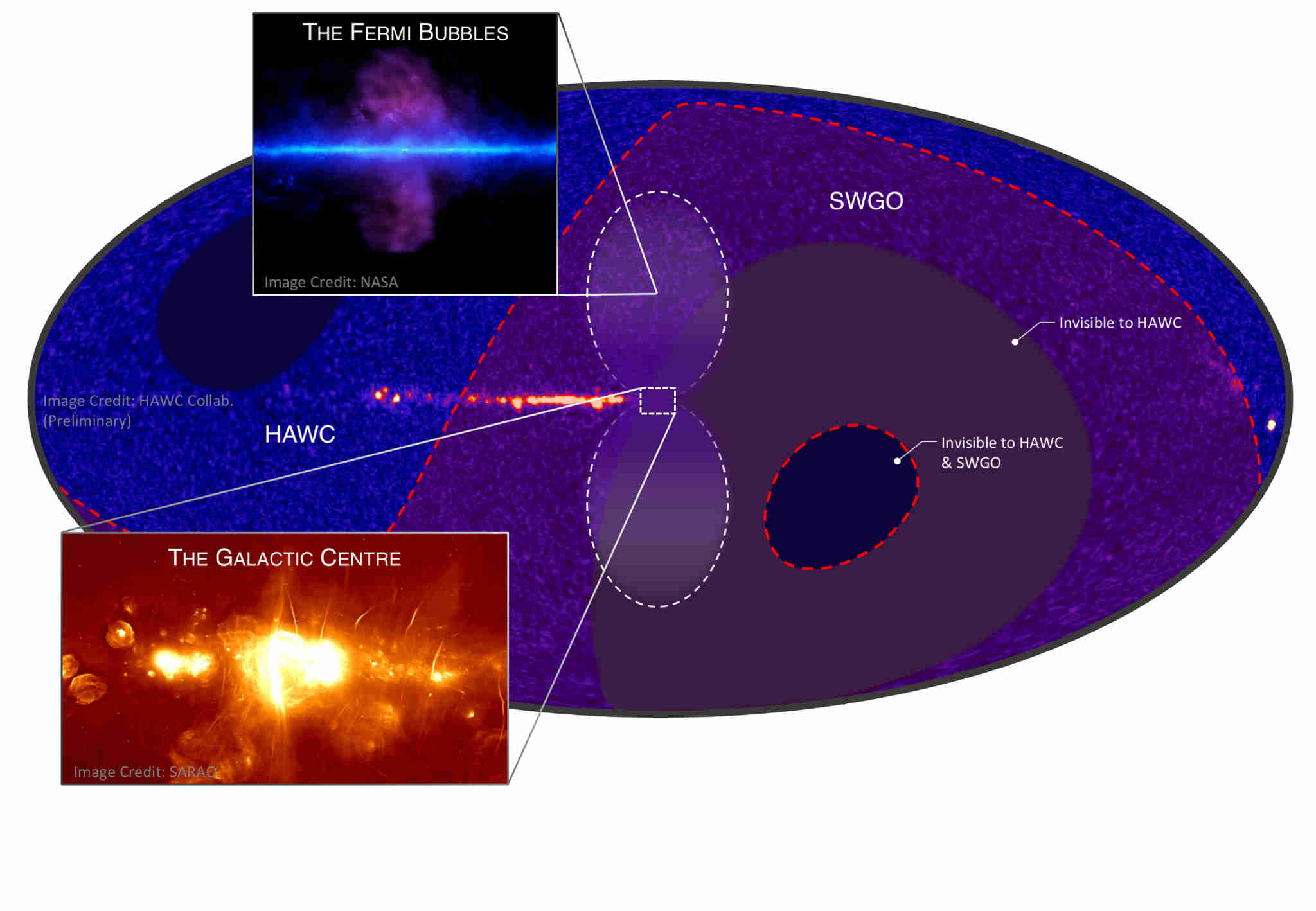}
    }
    \caption[SWGO]{SWGO design.}
    \label{fig:intro}
\end{figure}

The Southern Wide-field Gamma ray Observatory (SWGO) is an international collaboration with the objective of realizing a new-generation observatory for Very-High-Energy (VHE) gamma rays. The observatory will be located in South America at a latitude between -30° and -10°, at an altitude of at least 4.4 km. Some similar facilities already exist, such as HAWC in Mexico\cite{hawc_tev} and LHAASO in China\cite{LHAASO_PeV}, but SWGO will be the first observatory of its kind in the Southern hemisphere, which has a privileged view of the galactic center region\cite{loi,jim_conf,white_pap} (\cref{fig:fov}).

The observatory will be based on thousends of Water Cherenkov units (\cref{fig:array1}). In the design of such units, there are many variables to assess, such as the detector’s geometry, the types of sensors to be used and their configuration, and the materials; therefore some test facilities are needed\cite{water_fra}.

\section{The prototype tank}
\label{sec:prot_tank}%
\begin{figure}[h!]
    \centering
    \subfloat[B6 building in Politecnico di Milano's Bovisa campus: site of the test installation and first candidate for the tank's location.\label{fig:b6}]{
        \includegraphics[height=5.5cm]{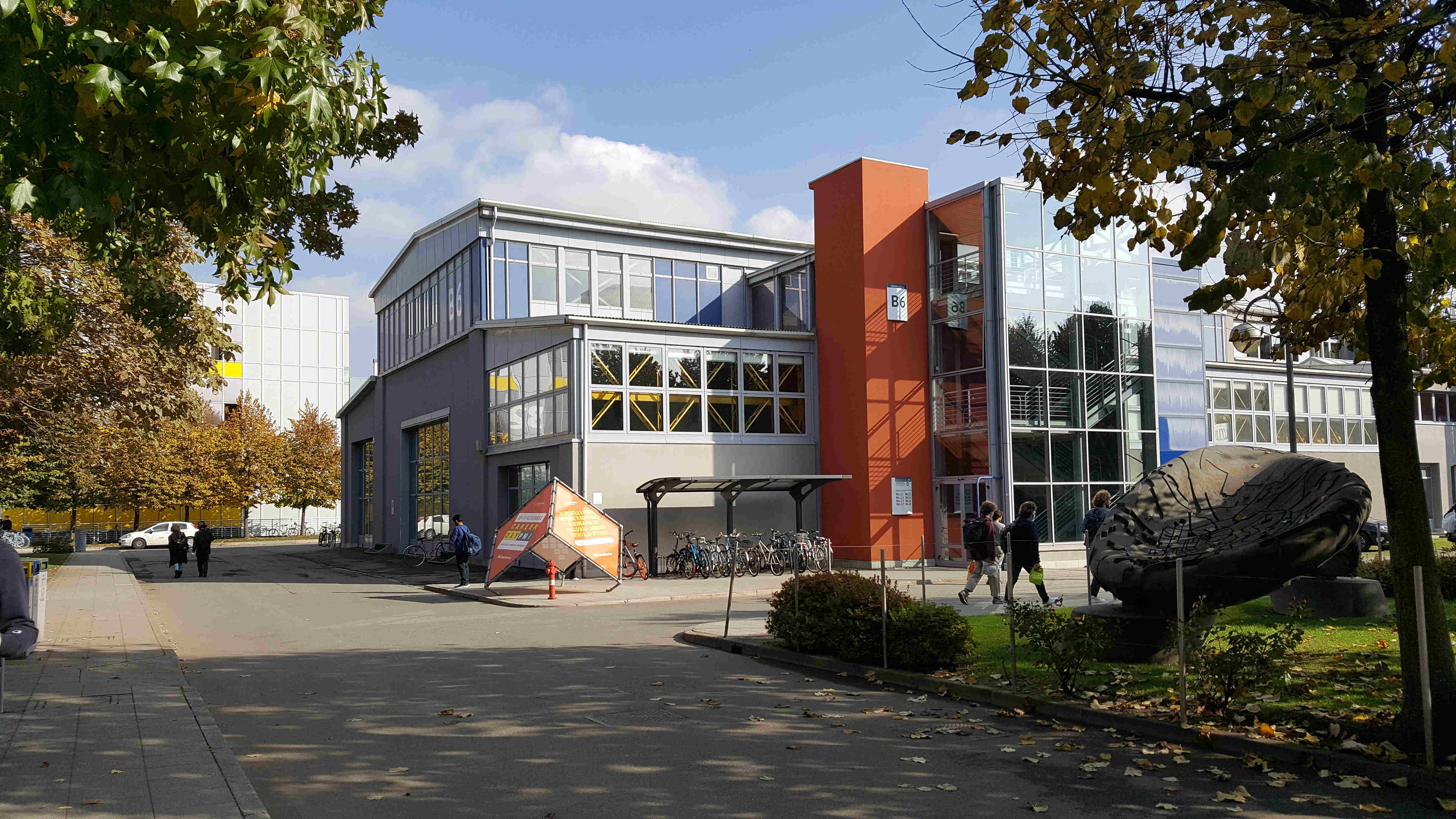}
    }
    \quad
    \subfloat[The prototype tank inside B6 labs after test installation.\label{fig:tank1}]{
        \includegraphics[height=5.5cm]{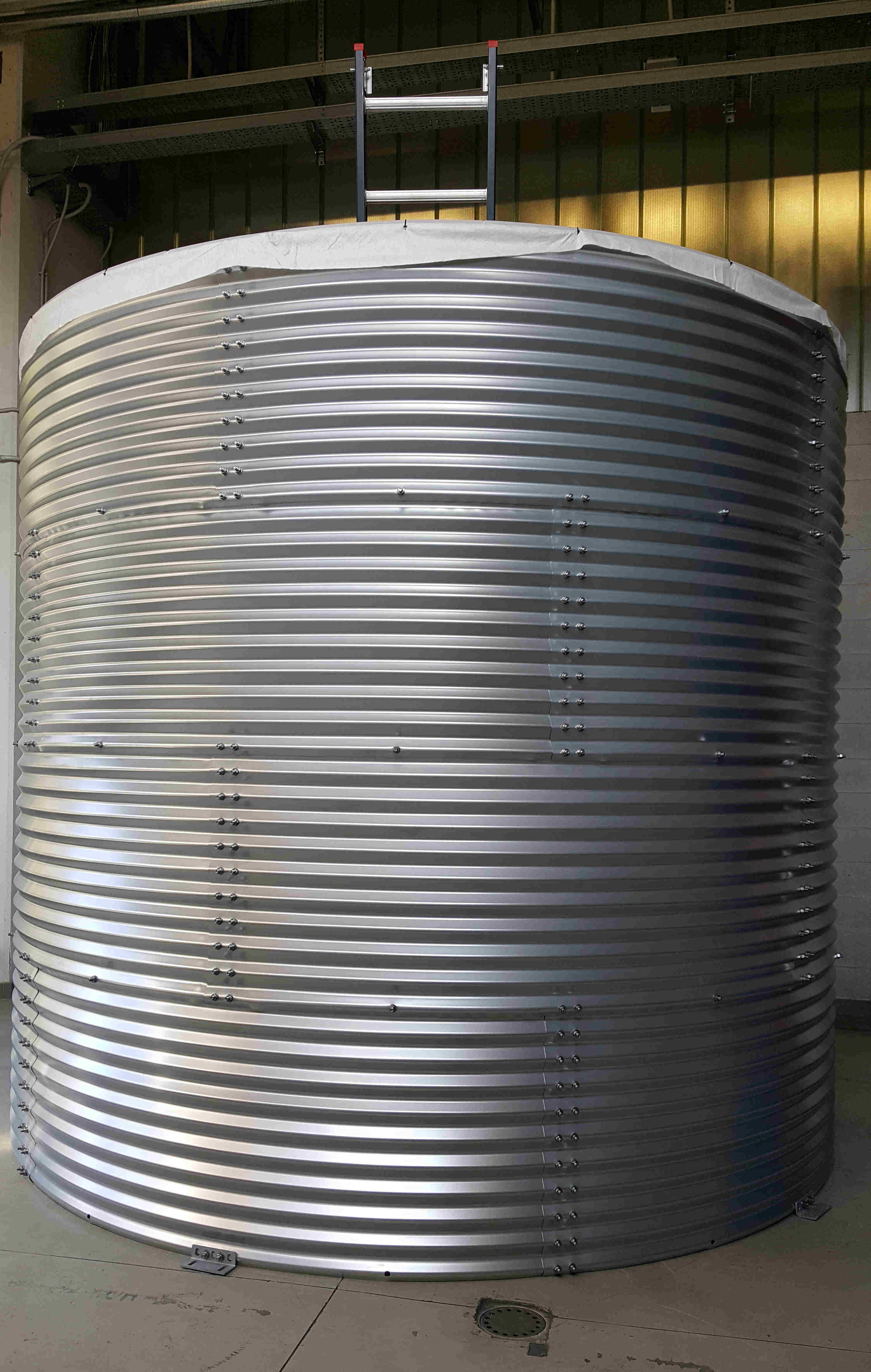}
    }
    \caption[Test installation of the Prototype Tank.]{Test installation of the Prototype Tank.}
    \label{fig:tank_location}
\end{figure}

The Italian partners of the SWGO collaboration (Istituto Nazionale di Fisica Nucleare, Politecnico di Milano, Università degli Studi di Torino, Università degli Studi di Padova and Università degli Studi di Napoli Federico II) are working on the realization of a test facility to be used for SWGO. It will consist of a cylindrical tank (3.32 m diameter and 3.12 m height) with an external structure made of galvanized steel and an internal liner in AQUATEX® PVC which could be then changed for different tests (\cref{fig:tank_location}). The prototype tank is installed at Politecnico di Milano, and its objective is to act as a test facility for different types of sensors and sensors' configurations, different types of liners, different tank configurations and anything else that could be useful to the collaboration.

Multi-PMT modules and light traps for example are being studied at the universities of Padova and Napoli respectively and the prototype tank will be used to test them. Multi-PMT modules are made by 3’’ PMTs, like KM3 Hyper-Kamiokade (\cref{fig:multipmt}), which has been considered as inspiration although a MoU is still being finalized. Light traps on the other hand use Wavelength Shiftters (WLS), which are materials that absorb light and re-emit it at a lower energy in a different random direction, so the light ends up trapped inside because of internal reflection (\cref{fig:wls}). This means that a small sensor can be used to capture the light by using a large WLS surface\cite{light_traps,wls}.

\begin{figure}[h!]
    \centering
    \subfloat[Hyper-K prototype considered as inspiration.\label{fig:multipmt}]{
        \includegraphics[height=5.1cm]{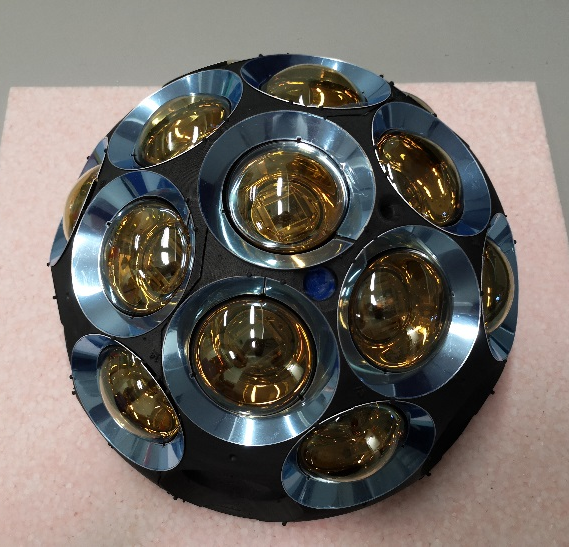}
    }
    \quad
    \subfloat[WLS light traps: concept and tests.\label{fig:wls}]{
        \includegraphics[height=5.1cm]{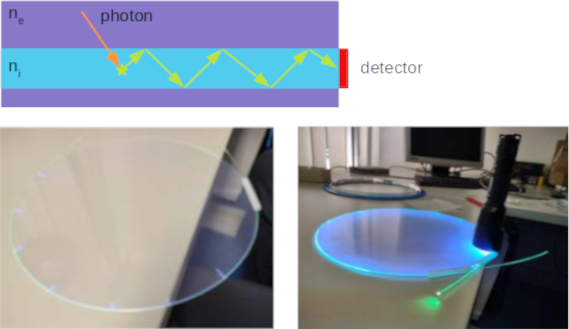}
    }
    \caption[Experimental sensors design.]{Experimental sensors design.}
    \label{fig:experimental}
\end{figure}

\subsection{Main requirements}
The first main requirement to be considered for the tank’s installation site is that the floor must resist the tank’s pressure due to the large volume of water contained. An analysis on detection capabilities as a function of the water level has therefore been done first, in order to choose the appropriate water level (see \cref{sec:wl}). The analysis's results could also be useful in the future for the studies on tank geometry.

Since the prototype tank is a test facility, it will be of main importance to be able to change the sensors’ configurations frequently, so an appropriate structure was designed to hold the sensors. During the design process, another main requirement that had to be taken into account was the need of maintaining high water purity inside the tank, which posed limits on the materials’ choices.

\section{Study on particle detection as a function of the water level}
\label{sec:wl}%

\begin{figure}[h!]
    \centering
    \subfloat[Prototype tank with 2.0 m water level.\label{fig:multipmt}]{
        \includegraphics[height=4.5cm]{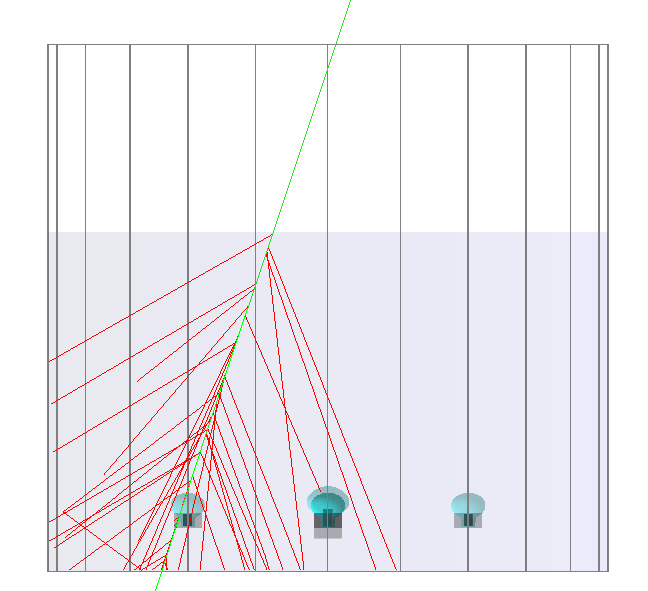}
    }
    \quad
    \subfloat[Prototype tank with 2.7 m water level.\label{fig:wls}]{
        \includegraphics[height=4.5cm]{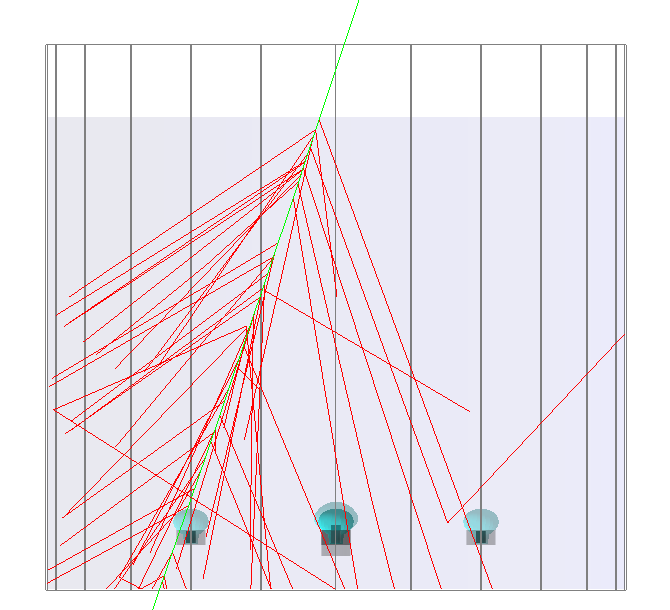}
    }
    \caption[water levels]{Visual output of the simulation of a muon (green) entering the tank and producing photons (red) in two different scenarios.}
    \label{fig:water_levels}
\end{figure}

A preliminary study has been conducted on two showers of muons with the energies of 1 GeV and 10 GeV: muons in fact are the only particles that will most likely be detected at Milano’s altitude. The shower’s interaction with the tank has been studied with 5 different water levels: 1.65 m, 2.00 m, 2.35 m, 2.70 m and 3.05 m (\cref{fig:water_levels}).

An additional study has been carried out on electron showers with the energy of 1 GeV to verify the model, since a different behavior was expected from the tank’s interaction with muons and electrons. Muons are more penetrating particles with respect to electrons: the latter lose energy faster, thus producing Cherenkov light only during a brief segment in the upper part of the detector; muons on the other hand usually cross the whole tank. A lower number of PE detected by the PMTs can be expected for electrons with a higher water level since they are produced only in the upper part of the tank and many of them would therefore be absorbed by the water before reaching the photosensors.

Both studies have been carried out using the \texttt{HAWCSim} framework, which makes use of \texttt{Geant4}\cite{geant4} to simulate the interaction of the particle with the tank itself and the water, including the production of Cherenkov photons that can be detected by the Photo Multiplier Tubes (PMTs) installed inside the tank (\cref{fig:water_levels}).

The tank’s dimensions and materials used for the simulations were the ones reported in \cref{sec:prot_tank}. The sensors’ configuration chosen (which will be the prototype tank's reference configuration) was made by:
\begin{itemize}
    \item $1\times 10$'' (253 mm) PMT situated in the center of the tank
    \item $4\times 5$'' (128 mm) PMT equidistant from the center in square configuration
\end{itemize}

In HAWCSim, three models of PMT from the Hamamatsu company were available at the time of this analysis: 8” R5912 PMT, 10” R7081HQE PMT and 3” R12199 PMT. To simulate the four 5” PMTs, the 8” R5912 PMTs were used and then scaled to 5’’ during the analysis phase\cite{water_fra}.

\subsection{Particles generation}
12000 particles (for each case: 1 GeV muons, 10 GeV muons and 1 GeV electrons) have been generated in a disk of fixed radius over the tank (r = 1.78 m, 10 cm larger than the tank’s radius; h = 3.22 m, 10 cm larger than the tank’s height). Their azimuth angle $\phi$ has been randomly selected in the range $\theta$ - 360° and zenith angle $\theta$ extracted from the distribution $\cos(\theta)^2$. In each scenario, the first 10000 particles entering the tank have been analyzed. Since they had random direction in fact, not all the particles would enter the tank, therefore generating 12000 particles assured having at least 10000 that could be used for the analysis\cite{fra_artile}.

\subsection{Results}
\begin{figure}[h!]
    \centering
    \subfloat[Number of PE detected. \label{fig:npe}]{
        \includegraphics[height=4.2cm]{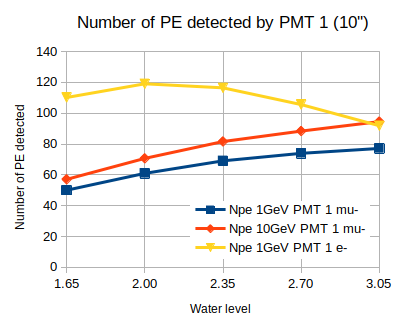}
    }
    \quad
    \subfloat[Detection efficiency. \label{fig:eff}]{
        \includegraphics[height=4.2cm]{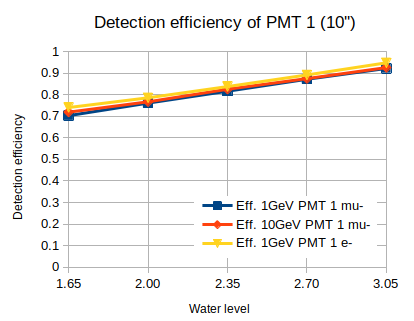}
    }
    \quad
    \subfloat[Standard deviation of the first photon time. \label{fig:std}]{
        \includegraphics[height=4.2cm]{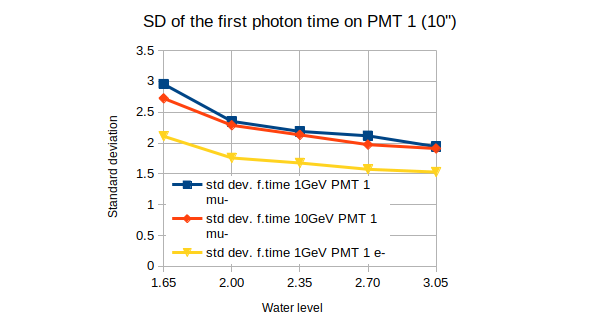}
    }
    \caption[Plots of results]{Plots of results for muons (1 GeV and 10 GeV) and electrons (1 GeV).}
    \label{fig:muons_gen}
\end{figure}

As it can be seen from \cref{fig:eff}, the detection efficiency increases linearly with the water level and the standard deviation of the first photon time decreases with the water level (\cref{fig:std}), meaning that the detector becomes more sensitive with more water. The number of photoelectrons detected is important to notice because here the difference between muons and electrons can be clearly appreciated (\cref{fig:npe}). The number of PE detected decreasing with higher levels of water for electrons was the expected behavior, in contrast with muons, for which that number increases, so the model was considered to be reliable.

\section{Design of the PMT holder}
The material chosen for the PMT holder has been stainless steel AISI 304, which had good mechanical properties and was compatible with purified water.

\begin{figure}[h!]
    \centering
    \subfloat[Initial structure scheme. \label{fig:config}]{
        \includegraphics[height=5.6cm]{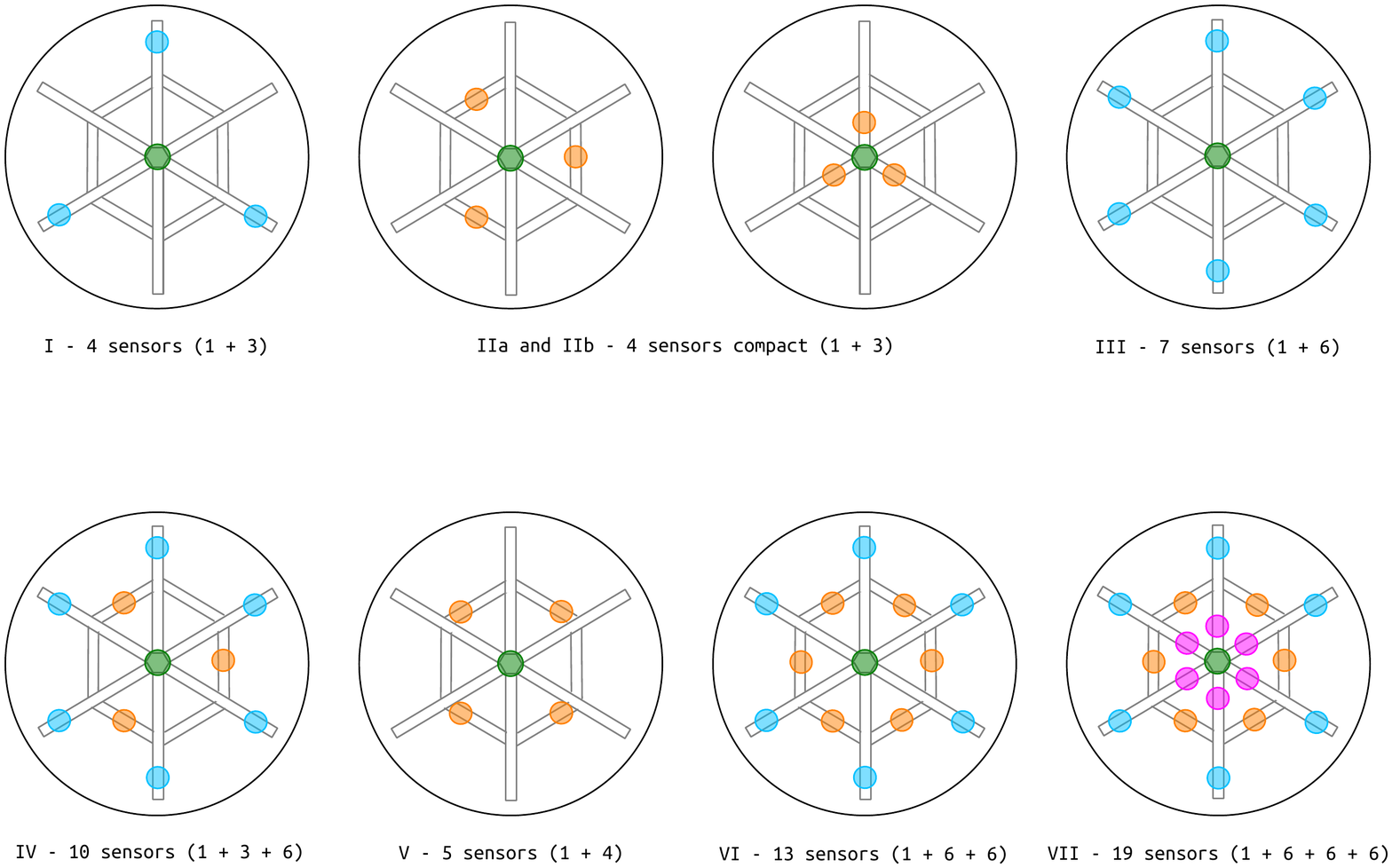}
    }
    \quad
    \subfloat[Initial structure cad model. \label{fig:bigcad}]{
        \includegraphics[height=3.8cm]{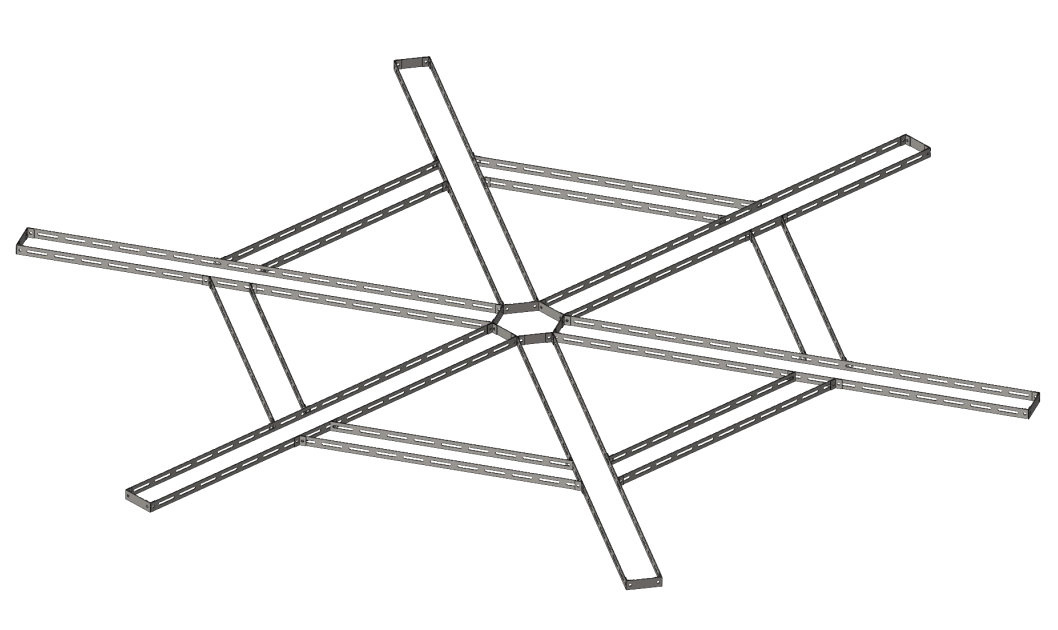}
    }
    \caption[First design of the PMT holder.]{First design of the PMT holder.}
    \label{fig:big}
\end{figure}

The first idea was to design a structure that could hold as many sensors configurations as possible. A CAD model of such structure had been realized (\cref{fig:big}), and it was a good compromise between weight, robustness and flexibility requirements. This design however had to be adapted to parts available on the market and it was not easy to find similar pieces. In addition to that, the PMTs of the reference configuration were nearly ready to be tested, while the other sensors that have to be tested here are still in development.

\begin{figure}[h!]
    \centering
    \includegraphics[width=0.58\textwidth]{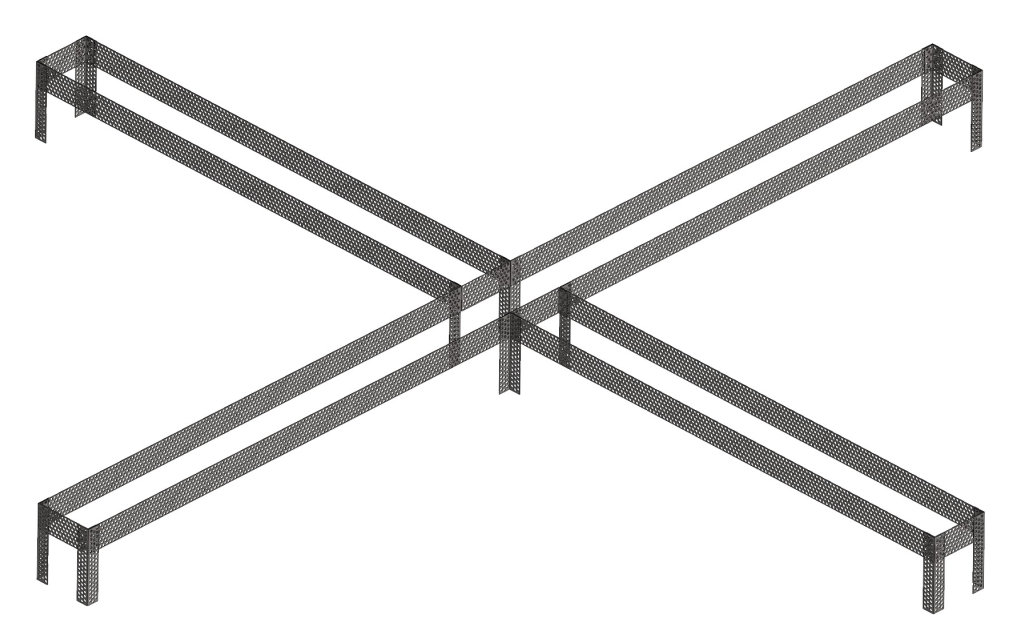}
    \caption{3D model of the cross holder.}
    \label{fig:cross_model}
\end{figure}

\begin{figure}[h!]
    \centering
    \subfloat[Part a: flat bar. \label{fig:bar}]{
        \includegraphics[height=3cm]{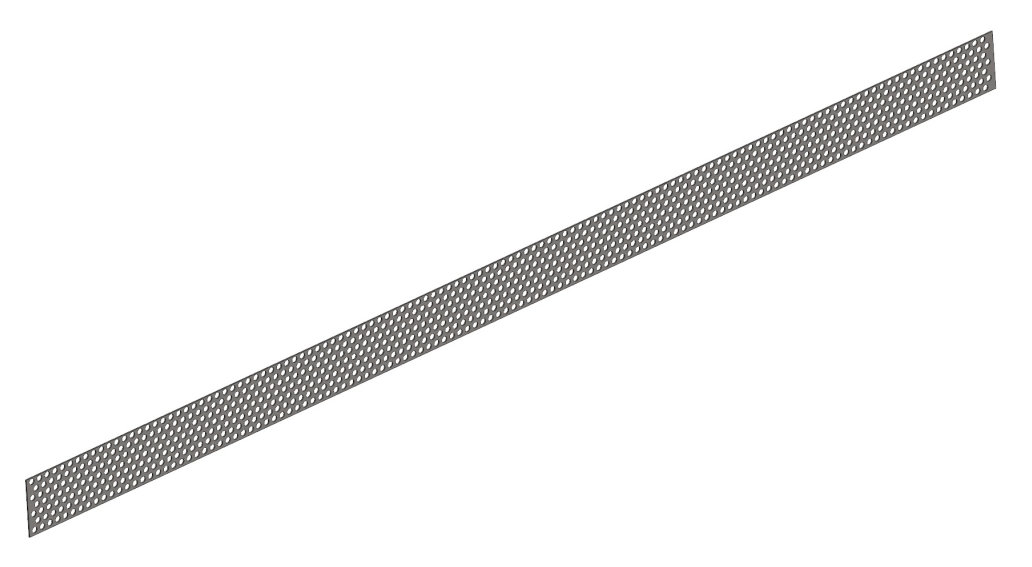}
    }
    \quad
    \subfloat[Part b: C-profile. \label{fig:dist}]{
        \includegraphics[height=3cm]{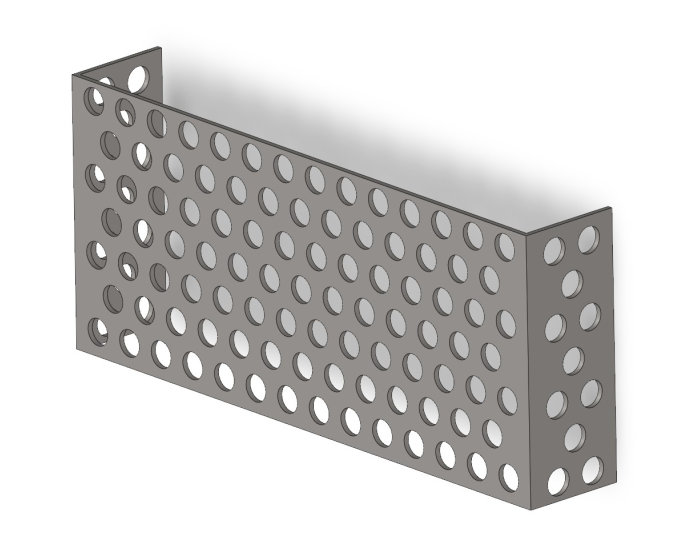}
    }
    \quad
    \subfloat[Part c: L-profile. \label{fig:ang}]{
        \includegraphics[height=3cm]{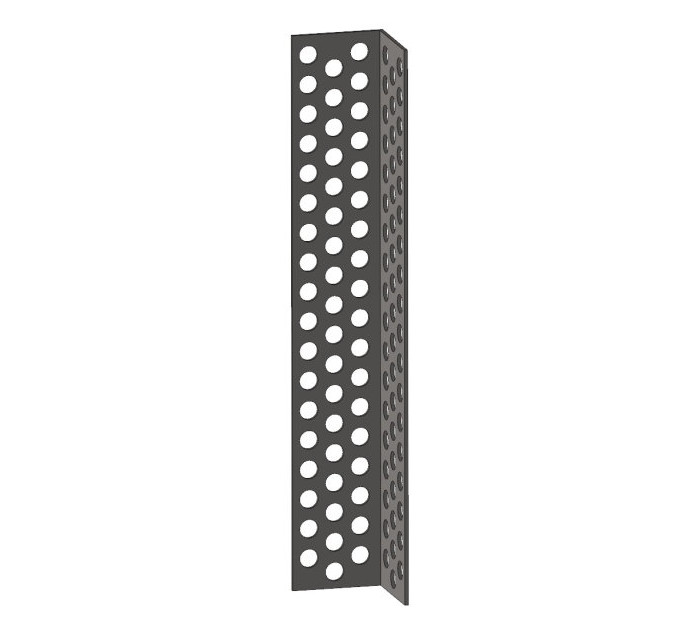}
    }
    \caption[Cross holder parts]{Parts composing the cross holder: part a is the base piece bought from the supplier (1000 mm x 50 mm x 1 mm perforated flat bar) from which part b and part c can be obtained.}
    \label{fig:parts}
\end{figure}

It has therefore been decided to design a simpler structure to be used for the reference configuration, designing it directly with parts that were known to be available on the market, with the idea of disassembling it and reuse the parts for the bigger structure when it will be needed (\cref{fig:cross_model} and \cref{fig:parts}).

\begin{figure}[h!]
    \centering
    \subfloat[Von Mises stress with static maximum load. \label{fig:cross_VM}]{
        \includegraphics[height=5cm]{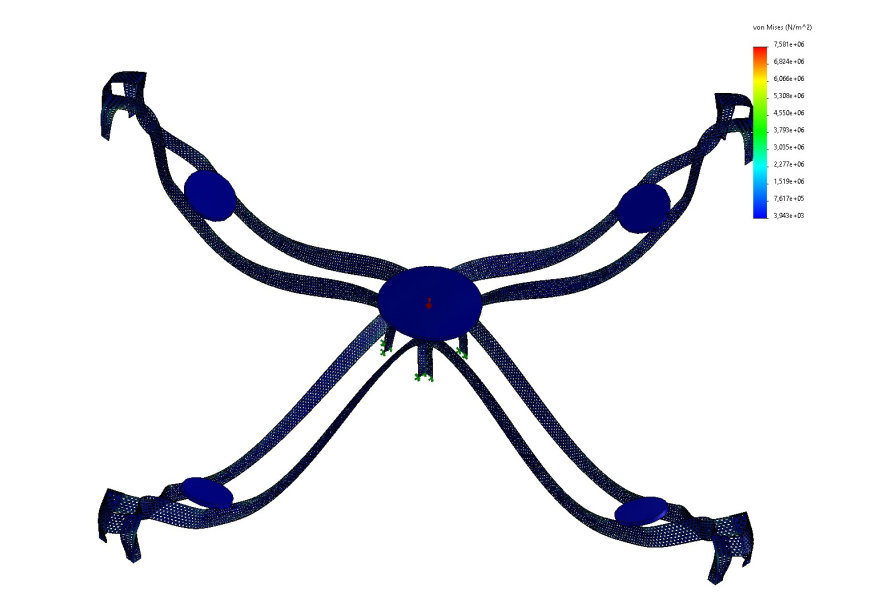}
    }
    \quad
    \subfloat[Displacement with static maximum load. \label{fig:cross_disp}]{
        \includegraphics[height=5cm]{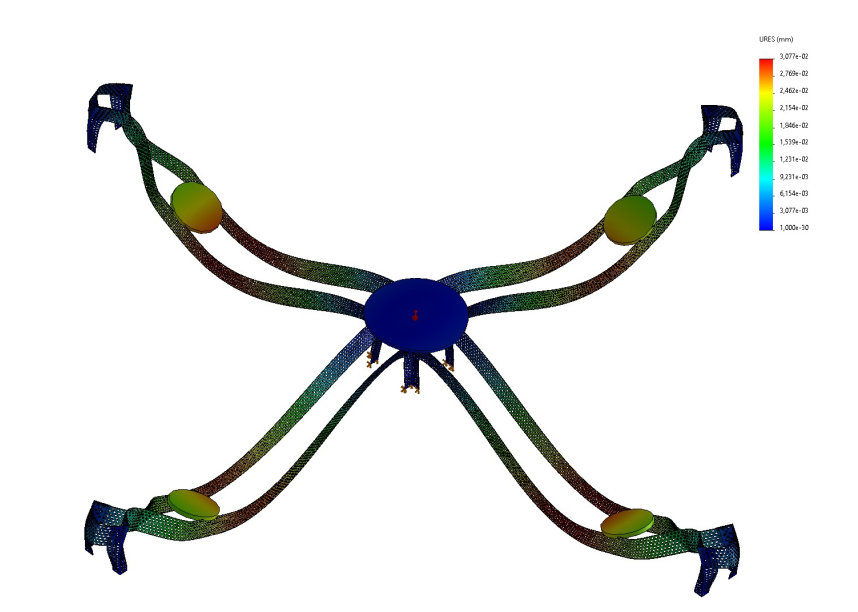}
    }
    \caption[Cross holder analysis]{Cross holder: results of the \texttt{solidworks} static analysis.}
    \label{fig:cross_sym}
\end{figure}

A loading analysis has been performed in \texttt{solidworks} applying the mass of the reference configuration PMTs and it has been shown that the structure could easily sustain the weight (\cref{fig:cross_sym}).

To properly hold the PMTs, the same C-profiles used for the main structure have been used since the materials were already available and they were able to guarantee good stability. An additional gasket has to be placed between the C-profiles and the PMT in order to protect the sensor.

The parts have been manufactured in the workshop of the Department of Aerospace Science and Technology (Politecnico di Milano) by cutting and bending the flat bars. The C-profiles and the L-profiles could be obtained by cutting and folding the original perforated flat bars.

\begin{figure}[h!]
    \centering
    \includegraphics[width=0.5\textwidth]{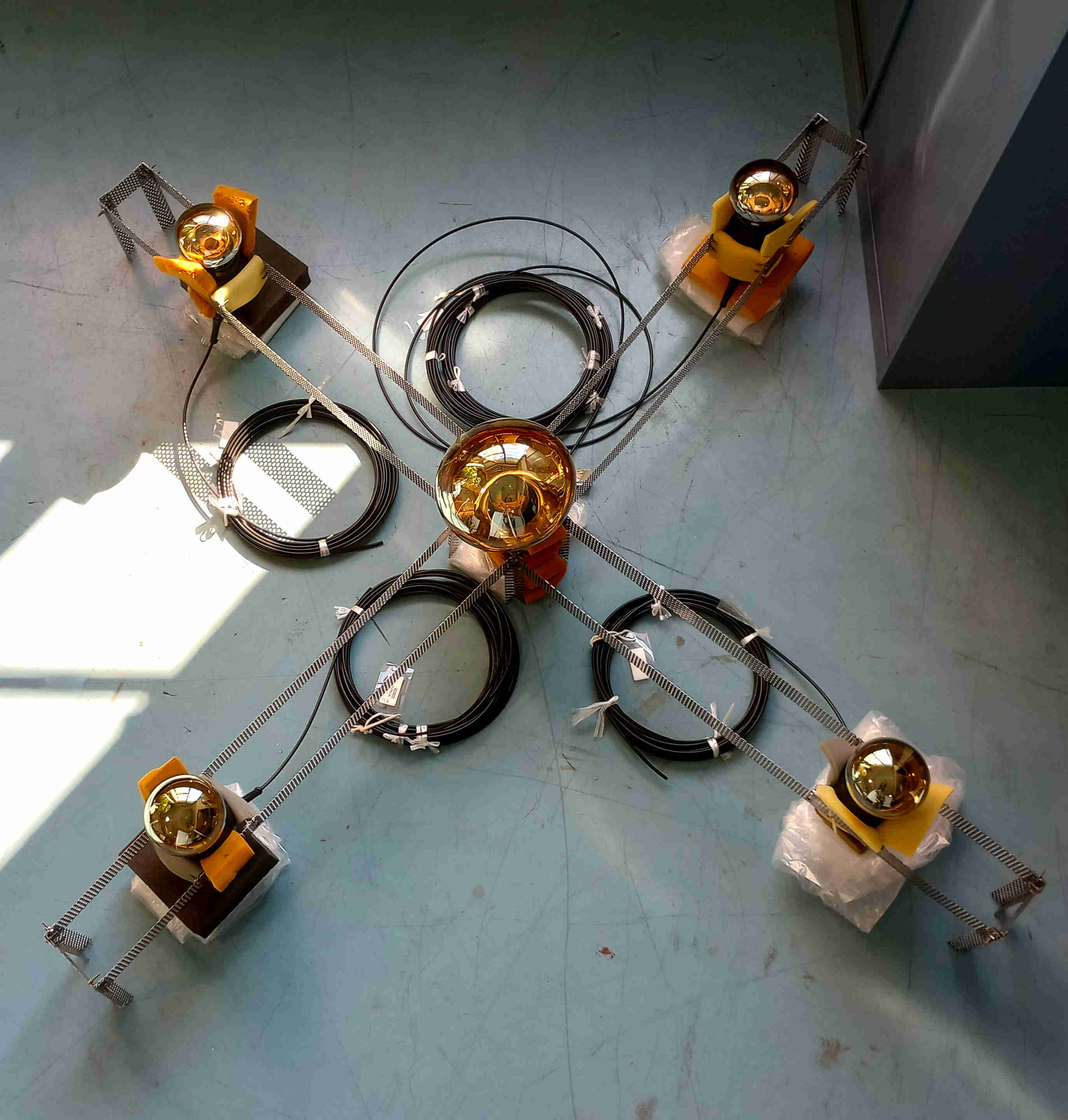}
    \caption{PMT holder: holding test.}
    \label{fig:holdtest}
\end{figure}

When the structure was complete, a placement test with the five reference configuration PMTs has been done. The appropriate rubber for the gaskets, compatible with purified water, had not been ordered yet, so foam rubber has been used. The structure was not deformed by the PMTs’ weight and the sensors were held stably (\cref{fig:holdtest}).

\section{Conclusions}
After considering the data coming from the study on particle detection as a function of the water level, it has been decided to maximize detection performances allowing the possibility to fill the tank completely.

Since the PMT holder is ready to be used, the first test with the reference configuration will start soon, followed by tests with the novel sensors designs from Napoli end Padova described in \cref{sec:prot_tank}.


\newpage

\begin{thebibliography}{99}

\bibitem{loi}
INFN and SWGO collaboration,
\emph{INFN SWGO letter of intent},
2020.

\bibitem{light_traps}
J.E. Ward, J. Cortina and D. Guberman,
\emph{Light-Trap: a SiPM upgrade for VHE astronomy and beyond},
\href{https://doi.org/10.1088/1748-0221/11/11/C11007}
{\emph{Journal of Instrumentation} \textbf{11} (2016)}.

\bibitem{wls}
M. Mariotti et al.,
\emph{Optimized wavelength shifters light traps with SiPM photo sensors for SWGO},
Internal SWGO paper.

\bibitem{jim_conf}
J. Hinton,
\emph{The Southern Wide-field Gamma-ray Observatory: status and srospects},
\href{https://doi.org/10.22323/1.395.0023}
{\emph{Proceedings of Science} \textbf{395 ICRC2021} (2022) 023}.


\bibitem{water_fra}
F. Bisconti and A. Chiavassa,
\emph{Study of water Cherenkov detector designs for the SWGO experiment},
\href{https://doi.org/10.22323/1.395.0895}
{\emph{Proceedings of Science} \textbf{395 ICRC2021} (2022) 895}.


\bibitem{geant4}
S. Agostinelli et al.,
\emph{GEANT4 — a simulation toolkit},
\href{https://doi.org/10.1016/S0168-9002(03)01368-8}
{\emph{Nuclear Instruments and Methods in Physics Research Section A: Accelerators, Spectrometers, Detectors and Associated Equipment} \textbf{506} (2003) 250}.


\bibitem{fra_artile}
F. Bisconti and A. Chiavassa,
\emph{Study of water Cherenkov detector design for ground-based gamma-ray experiments},
\href{https://doi.org/10.48550/arXiv.2205.02148}
{\emph{arXiv e-prints} (2022)}.


\bibitem{white_pap}
A. Albert et al.,
\emph{Science case for a wide field-of-view very-high-energy gamma-ray observatory in the Southern hemisphere},
\href{https://doi.org/10.48550/arXiv.1902.08429}
{\emph{arXiv e-prints} (2019)}.


\bibitem{LHAASO_PeV}
F. A. Aharonian,
\emph{LHAASO: A PeVatrons explorer},
\href{https://doi.org/10.1007/s11433-021-1751-8}
{\emph{Science China Physics, Mechanics \& Astronomy volume} \textbf{64} (2021) 109531}.


\bibitem{hawc_tev}
G. Sinnis,
\emph{The HAWC TeV gamma-ray observatory},
\href{https://doi.org/10.1393/ncc/i2011-10851-8}
{\emph{Il nuovo cimento C} \textbf{34} (2021) 65}.



\end{thebibliography}

\end{document}